\documentclass[prl,twoside,preprintnumbers,superscriptaddress,twocolumn,nofootinbib,nobibnotes]{revtex4}
\usepackage{hyperref}
\usepackage{amsmath,graphicx,slashed,color}
\usepackage{graphicx,epsfig,amssymb,bm}
\usepackage[utf8]{inputenc}
\def \beq{\begin{equation}}
\def \eeq{\end{equation}}
\def \beqa{\begin{eqnarray}}
\def \eeqa{\end{eqnarray}}

\usepackage{ragged2e}

\usepackage{booktabs}

\usepackage{tabularx}
\newcolumntype{C}[1]{>{\centering\arraybackslash}p{#1}}

\usepackage{graphicx}
\usepackage{url}
\usepackage{bm}
\usepackage{hyperref}
\usepackage{amsmath}
\usepackage{amssymb}
\usepackage{listings}
\usepackage{color}

\usepackage{tabularx}
\newcolumntype{C}[1]{>{\centering\arraybackslash}p{#1}}

\def\bea{\begin{eqnarray}}
\def\eea{\end{eqnarray}}

\def\gsim{\mathrel{\rlap{\lower4pt\hbox{\hskip1pt$\sim$}}
    \raise1pt\hbox{$>$}}}         
\def\lsim{\mathrel{\rlap{\lower4pt\hbox{\hskip1pt$\sim$}}
    \raise1pt\hbox{$<$}}}         

\newcommand{\be}{\begin{equation}}
\newcommand{\ee}{\end{equation}}
\newcommand{\bi}{\begin{itemize}}
\newcommand{\ei}{\end{itemize}}
\newcommand{\ben}{\begin{enumerate}}
\newcommand{\een}{\end{enumerate}}

\usepackage{xcolor}
\usepackage[utf8]{inputenc}
\graphicspath{{./figs/}}

\usepackage{amssymb}
\usepackage{dcolumn}	
\usepackage{bm}			
\usepackage{bbm} 		
\usepackage{multirow}
\usepackage{slashed}
\usepackage{blkarray}
\usepackage{booktabs}
\usepackage{makecell}
\usepackage{tikz}
\usepackage[compat=1.1.0]{tikz-feynman}
\usepackage{placeins}
\usepackage{listings}
\usepackage{amsmath}
\usepackage{xcolor}
\usepackage{pythonhighlight}
\usepackage{subcaption} 

\def\gsim{\mathrel{\rlap{\lower4pt\hbox{\hskip1pt$\sim$}}
    \raise1pt\hbox{$>$}}}         
\def\lsim{\mathrel{\rlap{\lower4pt\hbox{\hskip1pt$\sim$}}
    \raise1pt\hbox{$<$}}}         

\definecolor{codegreen}{rgb}{0,0.6,0}
\definecolor{codegray}{rgb}{0.5,0.5,0.5}
\definecolor{codepurple}{rgb}{0.58,0,0.82}
\definecolor{backcolour}{rgb}{0.95,0.95,0.92}

\lstdefinestyle{mystyle}{
    backgroundcolor=\color{backcolour},   
    commentstyle=\color{codegreen},
    keywordstyle=\color{magenta},
    numberstyle=\tiny\color{codegray},
    stringstyle=\color{codepurple},
    basicstyle=\ttfamily\footnotesize,
    breakatwhitespace=false,         
    breaklines=true,                 
    captionpos=b,                    
    keepspaces=true,                 
    numbers=left,                    
    numbersep=5pt,                  
    showspaces=false,                
    showstringspaces=false,
    showtabs=false,                  
    tabsize=2
}
\definecolor{Red}{rgb}{1.,0.,0.}

\newcommand{\mcO}{\mathcal{O}}

\usepackage[normalem]{ulem}
\definecolor{ao}{rgb}{0.0, 0.5, 0.0}
\definecolor{mt}{RGB}{108, 210, 92}

\newcommand{\bc}{\begin{center}}
\newcommand{\ec}{\end{center}}
\newcommand{\ba}{\begin{array}}
\newcommand{\ea}{\end{array}}

\usepackage{tabularx}

\usepackage[normalem]{ulem}
\definecolor{ao}{rgb}{0.0, 0.5, 0.0}
\definecolor{mt}{RGB}{108, 210, 92}

\begin{document}
\title{Higgs trilinear coupling in the standard model effective field theory at the high luminosity LHC and the FCC-ee}

\author{Jaco~ter~Hoeve}
\affiliation{The Higgs Centre for Theoretical Physics, University of Edinburgh, 
Mayfield Road, Edinburgh EH9 3JZ, United Kingdom}

\author{Luca~Mantani}
\affiliation{Instituto de Fisica Corpuscular (IFIC), Universidad de Valencia-CSIC, E-46980 Valencia, Spain}

\author{Juan Rojo}
\affiliation{Department of Physics and Astronomy, VU Amsterdam, De Boelelaan 1081, NL-1081, HV Amsterdam, The Netherlands}
\affiliation{Nikhef, Science Park 105, NL-1098 XG Amsterdam, The Netherlands}

\author{Alejo~N.~Rossia}
\affiliation{Dipartimento di Fisica e Astronomia “G. Galilei”, Universit\`a di Padova, and Istituto Nazionale
di Fisica Nucleare, Sezione di Padova, Via F. Marzolo 8, I-35131, Padova, Italy}

\author{Eleni~Vryonidou}
\affiliation{Department of Physics and Astronomy, University of Manchester, Oxford Road, Manchester M13
9PL, United Kingdom}

\date{\today}

\begin{abstract}
Motivated by the updated HL-LHC projections for Higgs pair production from  ATLAS and CMS and by the release of the FCC-ee Feasibility Study, we critically revisit the sensitivity of the global SMEFT analysis to deformations of the Higgs self-coupling modifier $\kappa_3$. 
To this end, we quantify the impact of SMEFT operators modifying double Higgs production at the LHC and single Higgs production, including loop corrections, at the FCC-ee, and include Renormalisation Group Evolution throughout.
We demonstrate that significantly improving on the legacy HL-LHC constraints on $\kappa_3$ at the FCC-ee is not possible without the $\sqrt{s}=365$ GeV run; that individual and
marginalised determinations are similar at the HL-LHC while differing by up to a factor
3 at the FCC-ee; and that quadratic EFT corrections cannot be neglected, { especially at the (HL-)LHC. This finding fundamentally alters the picture compared to strategic planning often based on one-parameter projections.}
Overall, the combination of HL-LHC and FCC-ee data offers unique potential to pin down the Higgs self-coupling with $\sim$$15\%$ precision.

\end{abstract}

\maketitle

\noindent
{\bf Introduction.}
The observation of Higgs pair production, and the associated constraints on the Higgs self-interactions, represent a cornerstone of the physics program of the HL-LHC~\cite{Cepeda:2019klc,Atlas:2025kye}.
Scrutinising the Higgs trilinear coupling is also a core goal of many proposed future particle colliders, from the FCC-ee~\cite{FCC:2018evy,Benedikt:2928193} and FCC-hh~\cite{FCC:2018vvp,Benedikt:2928193} to ILC/LCF~\cite{ILC:2013jhg,LinearColliderVision:2025hlt}, CEPC~\cite{CEPCPhysicsStudyGroup:2022uwl}, CLIC~\cite{Linssen:2012hp,Aicheler:2012bya}, and the Muon Collider~\cite{Accettura:2023ked,InternationalMuonCollider:2024jyv}. 

In the Standard Model (SM), the Higgs trilinear self-coupling is uniquely determined by the Higgs vacuum expectation value $(v)$ and its mass ($m_h$). 
Deviations from this prediction represent a smoking gun signal of the presence of new physics beyond the SM (BSM).
The precise determination of the Higgs self-coupling has also profound implications for the vacuum stability of the Universe~\cite{Hiller:2024zjp,Andreassen:2017rzq} and could offer novel insights towards a solution of the baryon asymmetry puzzle~\cite{Zhang:2023jvh,Huang:2016cjm,Bednyakov:2015sca}.

If deviations from the SM are present, the Higgs potential can be parameterised as
\be
\nonumber
V(h) = \frac{1}{2} m_h^2 h^2 + \lambda_{\rm SM}(1 + \delta \kappa_3) v h^3 
+ \frac{1}{4} \lambda_{\rm SM}(1 + \delta \kappa_4) h^4 \, ,
\ee
where $\lambda_{\rm SM} = m_h^2/2v^2$.  In the SMEFT framework~\cite{Isidori:2023pyp}, the leading deviation in the Higgs trilinear coupling stems from the purely-Higgs dimension-six operator  
\be
\label{eq:operator_purelyhiggs}
\mathcal{O}_\varphi = \left( \varphi^\dagger \varphi - \frac{v^2}{2} \right)^3 
\supset v^3 h^3 + \frac{3}{2} v^2 h^4 \, ,
\ee
governed by the Wilson coefficient $c_\varphi$. 
Additionally, in the Warsaw basis, two more operators ($c_{\varphi \Box}$ and $c_{\varphi D}$) modify the Higgs potential via field redefinitions~\cite{Alasfar:2023xpc}.
Hence in the dimension-six SMEFT, the Higgs trilinear coupling shifts from its SM value ($\delta\kappa_3=0$) as 
\be
\label{eq:kappa3_warsaw}
\delta \kappa_3
= -\frac{2v^4}{m_h^2} \frac{c_\varphi}{\Lambda^2} 
+ \frac{3 v^2}{\Lambda^2} \left( c_{\varphi \Box} - \frac{1}{4} c_{\varphi D} \right) \, ,
\ee
with furthermore $\delta \kappa_4 = \delta \kappa_3/6$.
By means of Eq.~(\ref{eq:kappa3_warsaw}), SMEFT results can be translated into constraints on the Higgs self-coupling via bounds on $\kappa_3 = 1 + \delta \kappa_3$.

Constraints on $\kappa_3$ at the LHC are provided primarily by Higgs pair production, for which the Run 2 bounds are $-1.2~(-1.2)\le \kappa_3 \le 7.2~(7.5)$ for ATLAS (CMS) at the 95\% C.L.~\cite{ATLAS:2024ish,CMS:2024awa}.
Subleading sensitivity is obtained through loop corrections to single Higgs processes~\cite{Gorbahn:2016uoy,Maltoni:2017ims,Degrassi:2016wml, Bizon:2016wgr}.
At future lepton colliders with $\sqrt{s} \lesssim 500$ GeV, such as the FCC-ee, CEPC, LEP3~\cite{Anastopoulos:2025jyh}, or the initial stage of the ILC/LCF, Higgs pair production remains kinematically inaccessible.
However, constraints on $\kappa_3$ can still be derived from loop-induced corrections to $Zh$ production~\cite{McCullough:2013rea,Maltoni:2018ttu,Asteriadis:2024xuk,Asteriadis:2024xts,Maura:2025rcv}.
Above the $Zhh$ production threshold, $\kappa_3$ could instead be accessed directly.

In the context of the ongoing European Strategy for Particle Physics Update (ESPPU26), the ATLAS and CMS collaborations have  presented new sensitivity projections for Higgs pair production at the HL-LHC~\cite{Atlas:2025kye}.
These projections indicate a combined significance of $7.6\sigma$ for di-Higgs production, corresponding to $\delta \kappa_3 =^{+29\%}_{-26\%}$ at the 68\% C.L., hence providing a substantial improvement in sensitivity compared to the previous (2019) projections~\cite{Cepeda:2019klc,deBlas:2019rxi} which reported $\delta \kappa_3 =^{+50\%}_{-48\%}$.
{Crucially,} these bounds are obtained from single-parameter analyses and neglect the correlations with EFT operators deforming both single and double Higgs production as well as other processes. {One of the aims of this work is to demonstrate explicitly how the situation changes fundamentally once correlations with the other EFT operators are properly accounted for.}
Another ESPPU26 milestone has been the completion of the FCC Feasibility Study~\cite{Benedikt:2928193}, which updates and improves previous projections~\cite{FCC:2018evy,2780507} and confirms four interaction points (IPs) as the baseline scenario for the FCC-ee. 

Motivated by these revised HL-LHC and FCC-ee projections, here we critically revisit the constraints on the Higgs self-coupling $\kappa_3$ within a state-of-the-art global SMEFT analysis~\cite{Celada:2024mcf,terHoeve:2025gey}.
We quantify the synergies between different input datasets: LHC Run 2, HL-LHC, FCC-ee@240GeV (always including the $\sqrt{s}=91$ and $165$ GeV runs), and FCC-ee (full). 
We determine the interplay between SMEFT operators deforming both single and double Higgs production, evaluate the convergence of the EFT expansion, and assess the impact of marginalisation as compared to one-parameter analyses.
We also examine the impact of these updated projections on UV completions of the SM that manifest primarily through modifications of the Higgs self-coupling, in particular for the custodial electroweak quadruplet model~\cite{Durieux:2022hbu}.

\vspace{0.1cm}
\noindent
{\bf Settings.}
Our analysis is based on the {\sc\small SMEFiT} framework~\cite{Hartland:2019bjb,Ethier:2021ydt, Giani:2023gfq,Ethier:2021bye}, in particular the {\sc\small SMEFiT3.0} release of~\cite{Celada:2024mcf} extended with RG evolution in~\cite{terHoeve:2025gey}, and whose matching to UV-complete models is described in~\cite{terHoeve:2023pvs}.
With respect to~\cite{terHoeve:2025gey}, our fitting basis is extended to $n_{\rm op}=56$ dimension-six operators with the addition of the two-quark-two-lepton operators $\mcO_{te}$, $\mcO_{t\ell}$, $\mcO_{Qe}$, $\mcO_{Q\ell}^{(-)}$, and $\mcO_{Q\ell}^{(3)}$ which are particularly relevant for $b\bar{b}$ and $t\bar{t}$ electroweak production.
Our input dataset follows~\cite{Celada:2024mcf} with the inclusion of the Run 2 constraints on $hh$ production from ATLAS~\cite{ATLAS:2024ish} and adding up to $n_{\rm dat} = 446$ observables and cross-sections from LEP (EWPO and diboson) and the LHC (Higgs, top, and diboson) assuming the SM.
The analysis is performed at both linear and quadratic order in the EFT expansion, incorporates NLO QCD corrections whenever available, and accounts for RGE evolution throughout. 

Projections for top, single Higgs, and diboson observables at the HL-LHC are extrapolated from Run 2 analyses with the method described in~\cite{Celada:2024mcf}.
The projections for double Higgs production follow the latest HL-LHC studies from ATLAS~\cite{ATL-PHYS-PUB-2025-006,ATL-PHYS-PUB-2024-016,ATL-PHYS-PUB-2025-001} and CMS~\cite{Collaboration:2928096} as well as their combination \cite{Atlas:2025kye}. 
We complement the projections for the inclusive signal strength $\mu_{hh}$~\cite{ATL-PHYS-PUB-2025-006} by incorporating differential information from the dominant $b\bar{b}\gamma\gamma$ and $b\bar{b}\tau^+\tau^-$ final states~\cite{ATL-PHYS-PUB-2024-016,ATL-PHYS-PUB-2025-001}, see the Supporting Information (SI) for further details.
We verify that the resulting di-Higgs likelihood provides a good approximation to the ATLAS/CMS combination by reproducing their sensitivity to $\kappa_3$ in a one-parameter fit.
The FCC-ee projections follow~\cite{Celada:2024mcf} based on four IPs~\cite{Benedikt:2928193}, and the associated theory predictions account for NLO electroweak corrections to the $Zh$ SMEFT cross-sections~\cite{Asteriadis:2024xts} at $\sqrt{s}=240$ and 365 GeV.
Theory uncertainties are neglected for the FCC-ee observables. 

In the following, results are presented for Wilson coefficients at a reference scale of $\mu_0 = 250$ GeV.
We report bounds on the EFT coefficients as Bayesian credible intervals (C.I.). 
While these coincide with confidence intervals in the Gaussian limit, they can differ significantly for non-Gaussian distributions.
This discrepancy is most pronounced in the (HL-)LHC fits, where C.I. tend to produce looser bounds compared to confidence intervals.

\begin{figure}[t]
    \centering
\makebox{\includegraphics[width=0.99\columnwidth]{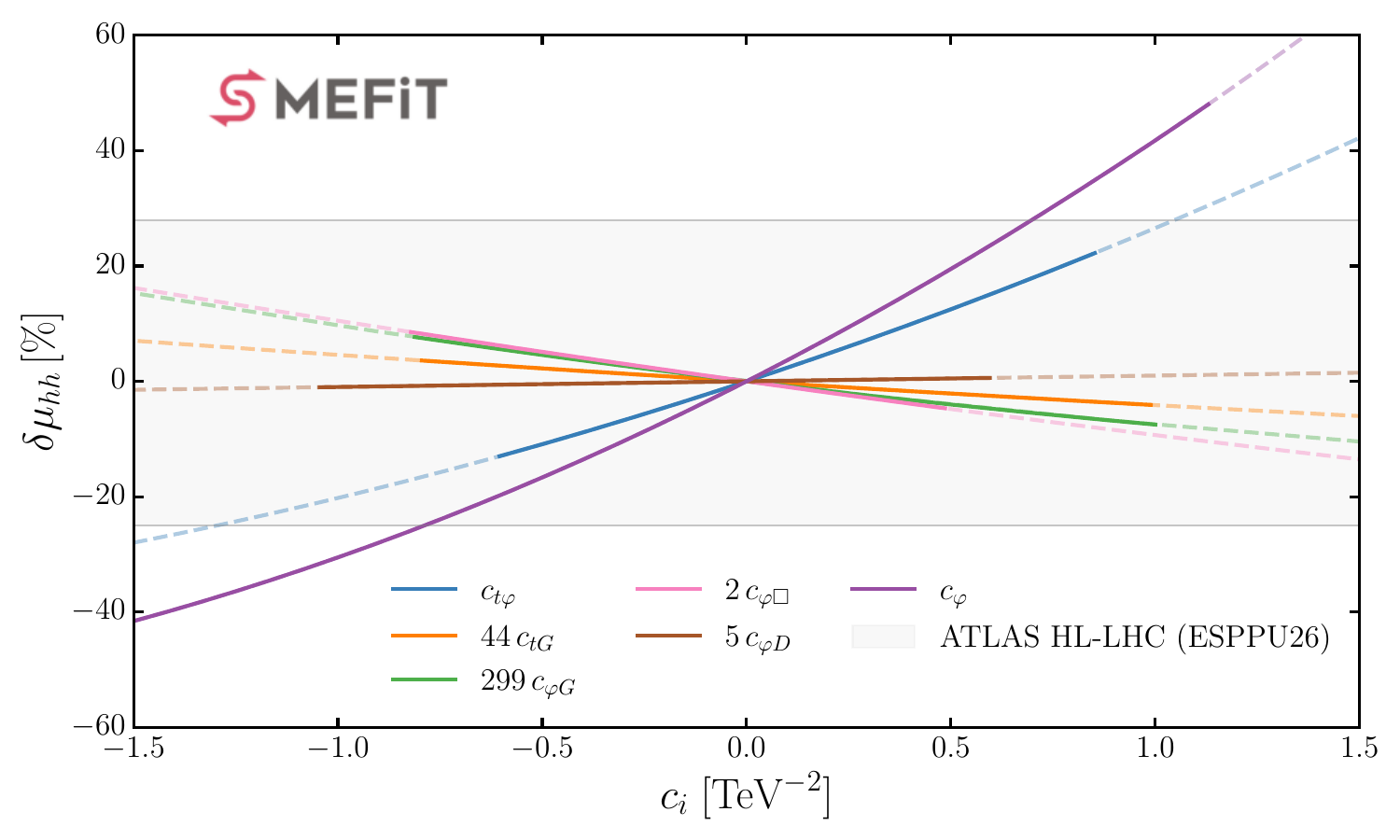}}\hfill
    \caption{
    Relative deformations (dashed) of the inclusive di-Higgs signal strength $\mu_{hh}$  for the relevant EFT coefficients $c_i$ and their 68\% C.I. bounds (solid lines) from the HL-LHC quadratic fit ($hh$ excluded).
    We also indicate the projected experimental bounds.
      }
\label{fig:HLLHC_HH_xs_marg_bounds}
\end{figure}

\vspace{0.1cm}
\noindent
{\bf The Higgs trilinear coupling at the HL-LHC.}
We consider first the global SMEFT fit at the HL-LHC based on the inputs described above.
To quantify the interplay between $c_\varphi$ and other operators entering $\mu_{hh}$ as well as single Higgs production, Fig.~\ref{fig:HLLHC_HH_xs_marg_bounds} displays the  relative deformations to $\mu_{hh}$ for each coefficient $c_i$ together with their 68\% C.I. from the HL-LHC quadratic fit with $hh$ projections excluded.
We also show the expected experimental bound on $\mu_{hh}$ from~\cite{ATL-PHYS-PUB-2025-006}.
Fig.~\ref{fig:HLLHC_HH_xs_marg_bounds} highlights that the sensitivity of di-Higgs production at the HL-LHC to operators other than $c_\varphi$ is marginal, as these are better constrained by other processes. Given this limited cross-talk, we anticipate that individual and marginalized bounds on $c_\varphi$ and $\delta\kappa_3$ will be similar at the HL-LHC.

\begin{figure}[t]
    \centering
    \includegraphics[width=0.99\linewidth]{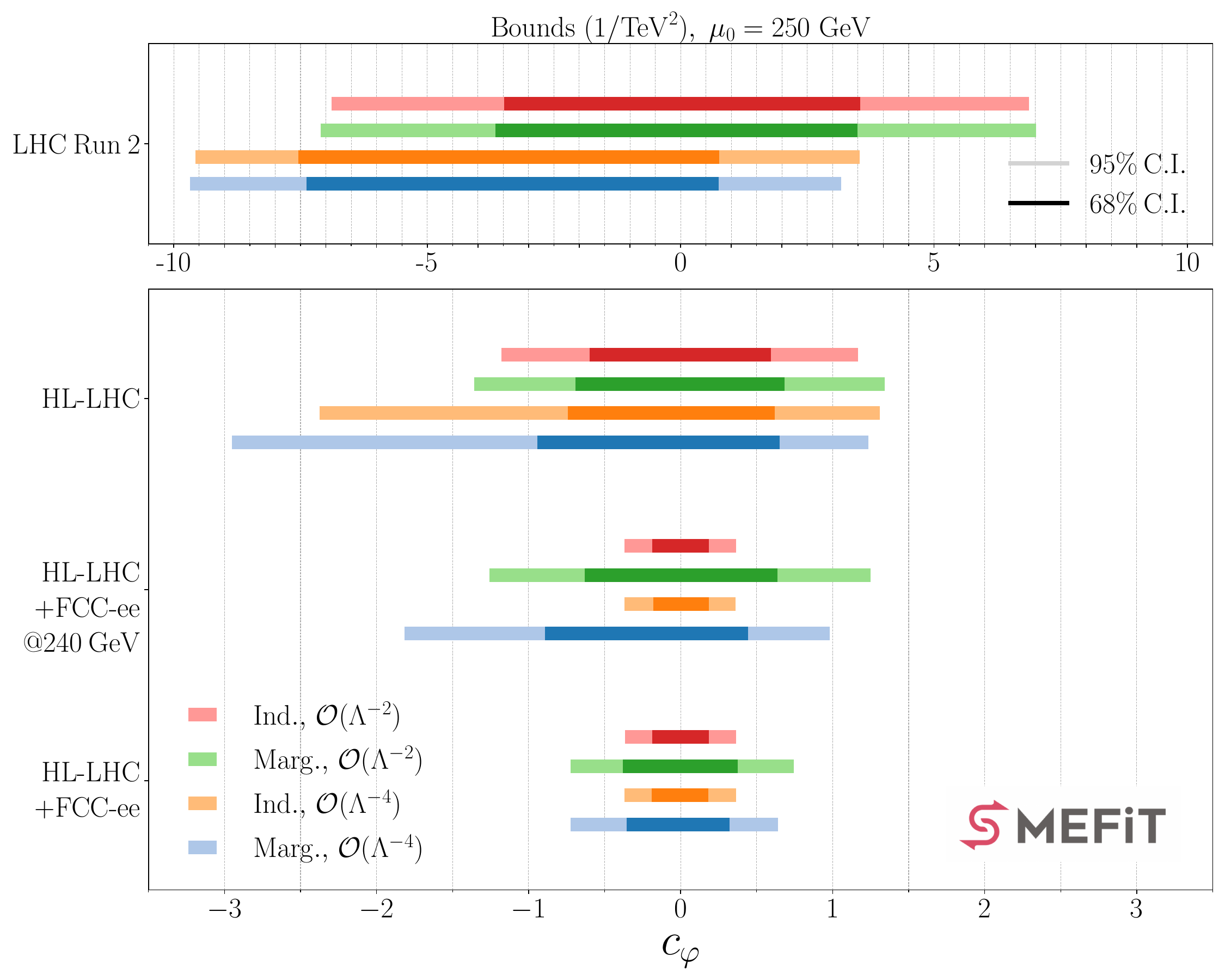}
    \includegraphics[width=0.99\linewidth]{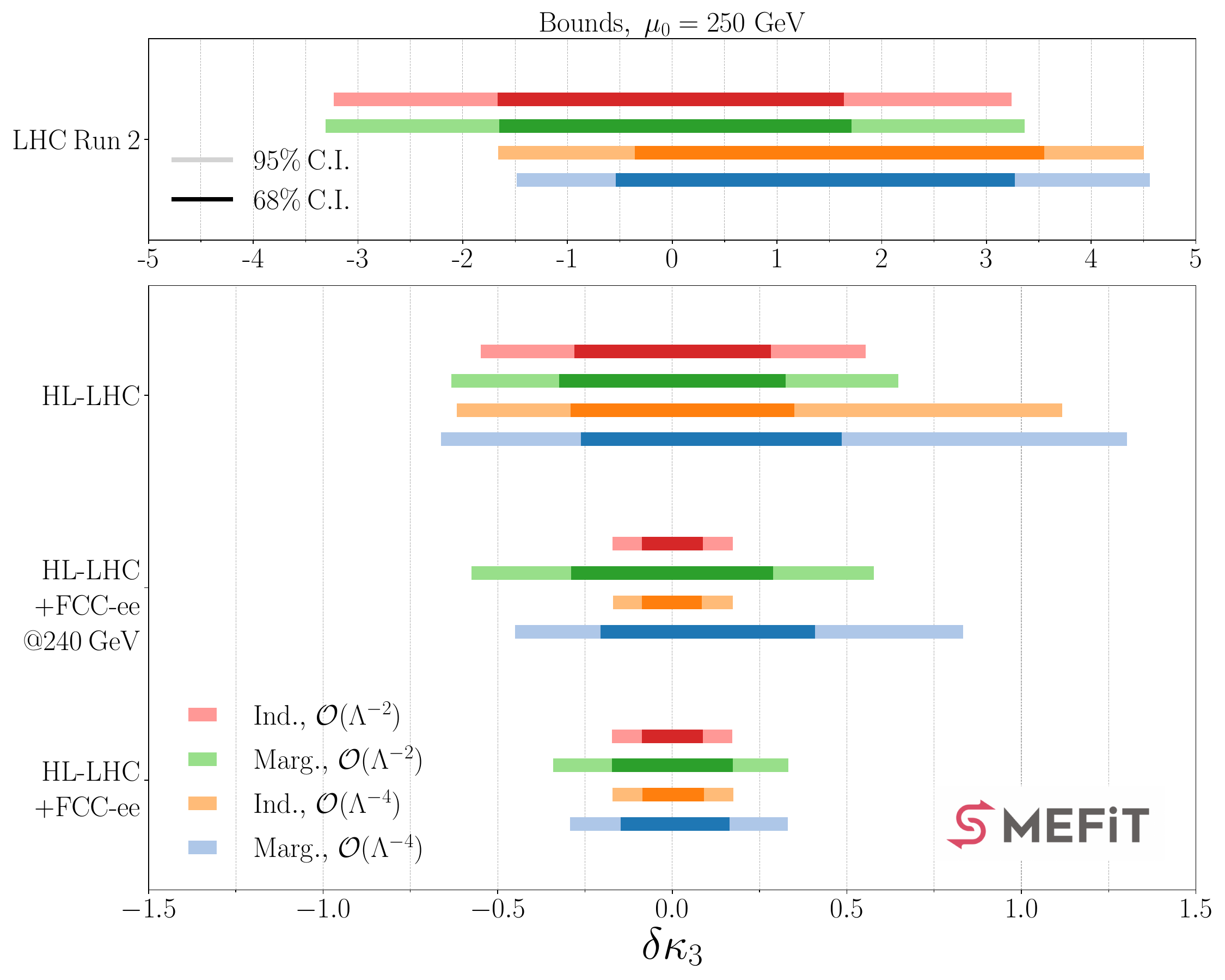}
    \caption{The 68\% (inner) and 95\%  (outer) C.I. bounds on $c_\varphi$ (upper) and $\delta\kappa_3$ (lower panel) in linear and quadratic EFT fits to different datasets. 
}
    \label{fig:bound_plot_cphi_hllhc_fcc}
\end{figure}

The upper (lower) panel of Fig.~\ref{fig:bound_plot_cphi_hllhc_fcc} displays the 68\% and 95\% C.I. bounds on $c_\varphi$ ($\delta\kappa_3)$ at $\mu_0=250$~GeV from individual and global marginalised fits, both at the linear and quadratic EFT levels, at the LHC Run 2 and at the HL-LHC, see Tables~\ref{tab:bounds_maintext} and~\ref{tab:bounds_all} for the numerical values.
From these results one finds that, first, the HL-LHC will improve the 68\% C.I. marginalised bounds on $\delta \kappa_3$ from Run 2 by a factor of 5~for both the linear and quadratic analyses.  
Second, individual and marginalised bounds are very similar, especially in the linear EFT case.
Indeed, the individual linear (quadratic) bound of $|\delta\kappa_3|\le 0.28$ ($-0.29 \le \delta \kappa_3 \le 0.35$) becomes $|\delta\kappa_3|\le 0.32$ 
($-0.26 \le \delta \kappa_3 \le 0.49$) in the marginalised case.
This result, consistent with Fig.~\ref{fig:HLLHC_HH_xs_marg_bounds}, indicates that in the global SMEFT analysis there is a negligible correlation between $\mathcal{O}_\varphi$ and other operators.
Third, quadratic EFT corrections to the determination of $\kappa_3$ at the HL-LHC are small but not negligible, loosening the linear bound by 15\%. 

\begin{table}[t]
    \renewcommand{\arraystretch}{1.40}
    \centering
    \begin{tabular}{|l|c|c|}
  \hline
         Input Dataset & $\quad$ EFT $\quad$ & \,\,\,{$\delta\kappa_3$ ($68\%$ C.I.)}\,\, \,\\
         \hline
         \multirow{2}{*}{LHC Run 2} & Linear & $[-1.68,\,1.68]$ \\
         \cline{2-3}
                  & Quad. & $[-0.54,\,3.27]$ \\
         \hline
                  \multirow{2}{*}{HL-LHC } & Linear & $[-0.32,\,0.32]$ \\
         \cline{2-3}
                  & Quad. & $[-0.26,\,0.49]$ \\
         \hline
         \multirow{2}{*}{HL-LHC\,\&\,FCC-ee$(240)$ } & Linear &  $[-0.29,\,0.29]$ \\
         \cline{2-3}
         \multirow{1}{*}{ }& Quad. & $[-0.20,\,0.41]$ \\
         \hline
         \multirow{2}{*}{HL-LHC\,\&\,FCC-ee}  & Linear &  $[-0.17,\,0.17]$ \\
         \cline{2-3}
         & Quad. &  $[-0.15,\,0.16]$\\
         \hline
    \end{tabular}
    \caption{Marginalised 68\% C.I. on $\delta\kappa_3$ (for $\mu_0=250$ GeV)
    in linear and quadratic global SMEFT fits.
 }
    \label{tab:bounds_maintext}
\end{table}

\begin{figure}[t]
    \centering
\makebox{\includegraphics[width=0.99\columnwidth]{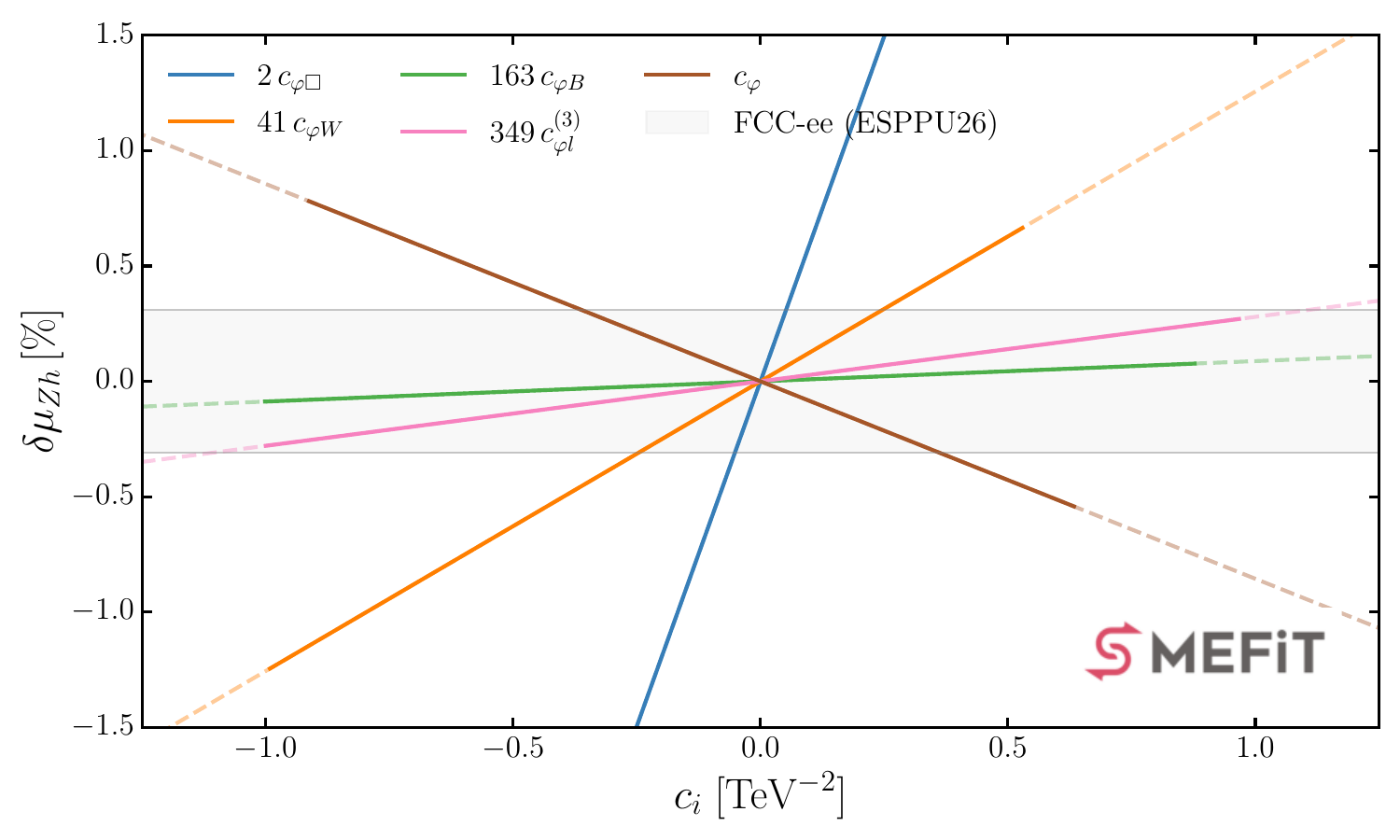}}\hfill
    \caption{
    Same as Fig.~\ref{fig:HLLHC_HH_xs_marg_bounds} for the inclusive  $\mu_{Zh}$ at the FCC-ee with $\sqrt{s}=240$ GeV, with the bounds from a HL-LHC+FCCee@240 quadratic fit (with $Zh$ excluded).
      }
\label{fig:FCC_ZH_xs_marg_bounds}
\end{figure}

\vspace{0.1cm}
\noindent
{\bf The Higgs trilinear coupling at the FCC-ee.}
As mentioned above,  $\mathcal{O}_\varphi$  can be constrained at the FCC-ee through NLO electroweak corrections to $Zh$ production~\cite{McCullough:2013rea,Asteriadis:2024xuk,Asteriadis:2024xts,Maura:2025rcv}.
To illustrate the correlation between $c_{\varphi}$ and other contributions to $Zh$ at the FCC-ee, Fig.~\ref{fig:FCC_ZH_xs_marg_bounds} presents the same information as Fig.~\ref{fig:HLLHC_HH_xs_marg_bounds} now for the projected $Zh$ signal strength measurement at $\sqrt{s} = 240$~GeV. 
The displayed 68\% C.I. constraints are derived from a global quadratic HL-LHC+FCCee@240 fit excluding $Zh$.
Unlike the HL-LHC case, at the FCC-ee several SMEFT operators will be constrained from $Zh$ at the same time as $c_{\varphi}$, in particular $c_{\varphi W}$ and $c_{\varphi \Box}$,  and hence one expects potentially large differences between individual and marginalised constraints for the Wilson coefficient $c_\varphi$.

These expectations are validated by the results shown in Fig.~\ref{fig:bound_plot_cphi_hllhc_fcc} and Tables~\ref{tab:bounds_maintext} and~\ref{tab:bounds_all}.
The marginalised 68\% C.I. bound on $\delta\kappa_3$ from the linear global SMEFT analysis at the FCC-ee is $|\delta\kappa_3| \leq 0.29$ without the $\sqrt{s} = 365~\text{GeV}$ run, and $|\delta\kappa_3| \leq 0.17$ when including it.  
Comparing with the corresponding projections for the HL-LHC, $|\delta\kappa_3|\le 0.32$, one finds that the FCC-ee operating at up to $\sqrt{s}=240$ GeV would not improve significantly on the HL-LHC legacy constraints, while the addition of the $\sqrt{s}=365$ GeV run enables a factor of two improvement.
A similar picture holds for the marginalised bound in the quadratic analysis, where the HL-LHC expectation of $-0.26 \le \delta\kappa_3 \le 0.49$ improves to  $-0.20 \le \delta\kappa_3 \le 0.41$ at FCC-ee@240 and $-0.15 \le \delta\kappa_3 \le 0.16$ after the full FCC-ee program.

A striking feature of Fig.~\ref{fig:bound_plot_cphi_hllhc_fcc}, consistent with Fig.~\ref{fig:FCC_ZH_xs_marg_bounds}, is the large difference between the individual and marginalised bounds on $\kappa_3$ at the FCC-ee@240. 
In the linear analysis, the individual bound of $|\delta\kappa_3|\le 0.09$ becomes a factor of 3 worse at the marginalised level, $|\delta\kappa_3|\le 0.29$, with a similar degradation found in the quadratic fits. 
This effect remains, albeit tamed, for the complete FCC-ee dataset, with individual and marginalised bounds differing by a factor of 2. 
We conclude that while di-Higgs production at the LHC  essentially corresponds to a direct measurement of $\kappa_3$, at the FCC-ee a global SMEFT interpretation is required to achieve robust constraints on the Higgs self-coupling from $Zh$ production. 

In the quadratic FCC-ee fit, $\mathcal{O}_{\varphi}$ is found to be primarily correlated with $\mathcal{O}_{\varphi \Box}$, with a correlation coefficient of $0.6$, suggesting that a more precise determination of $c_{\varphi \Box}$ could significantly reduce the gap between the marginalised and individual bounds. We also find that two-quark-two-lepton operators play an important role in the determination of $c_{\varphi}$, particularly before the $365~\mathrm{GeV}$ run is included. In their absence, the $68\%$ C.I. on $\kappa_3$ would already be reduced to about $|\delta \kappa_3|\le 0.20$ even without the $365~\mathrm{GeV}$ data, while with the full FCC dataset we obtain $|\delta \kappa_3|\le 0.14$, {a result which further strengthens the case for the $\sqrt{s}=365~\mathrm{GeV}$ run.}

\vspace{0.1cm}
\noindent
{\bf The electroweak custodial model.}
Although many BSM scenarios induce large deformations of $\kappa_3$, it is challenging to obtain such effects without a comparable distortion in single-Higgs production~\cite{DiLuzio:2017tfn,Falkowski:2019tft,Durieux:2022hbu}.
One remarkable model which bypasses this limitation is the custodial symmetric $\Theta_{1}+\Theta_{3}$ model, where one extends the SM with scalar electroweak quadruplets $\Theta_{1}$ and $\Theta_{3}$, of hypercharges $1/2$ and $3/2$ respectively, to form a custodial bi-quadruplet~\cite{Durieux:2022hbu}. 
The interactions between the heavy scalars and the SM, besides the ones mandated by gauge symmetry, are described by
\begin{equation}
\label{eq:quadruplet_lagrangian}
\mathcal{L}_{\rm UV} \supset -\lambda_{\Theta}\,\varphi^* \varphi^* \left( \varepsilon \varphi \right)\Theta_{1} -\lambda_{\Theta} \varphi^* \varphi^* \varphi^* \Theta_{3}/\sqrt{3} + \text{h.c.}\,
\end{equation}
where $\varphi$ is the Higgs boson and $\Theta_{1}$ and $\Theta_{3}$ share the same mass, here assumed to be $m_{\rm UV}=4$~TeV.
Following tree-level matching of Eq.~(\ref{eq:quadruplet_lagrangian}) to the SMEFT at dimension-six, only the $\mcO_{\varphi}$ operator is generated~\cite{deBlas:2017xtg,Durieux:2022hbu}. 
Other operators are generated at one loop, with $\mcO_{\varphi \Box}$ and $\mcO_{t\varphi}$ having the largest matching coefficients among them.
Including only one of the quadruplets, e.g. $\Theta_{1}$, does not change the situation at tree level, but leads to the custodially-violating operator $\mcO_{\varphi D}$ being generated at one loop, with important phenomenological consequences~\cite{Durieux:2022hbu}.

\begin{figure}[t]
    \centering
\includegraphics[width=0.99\linewidth]{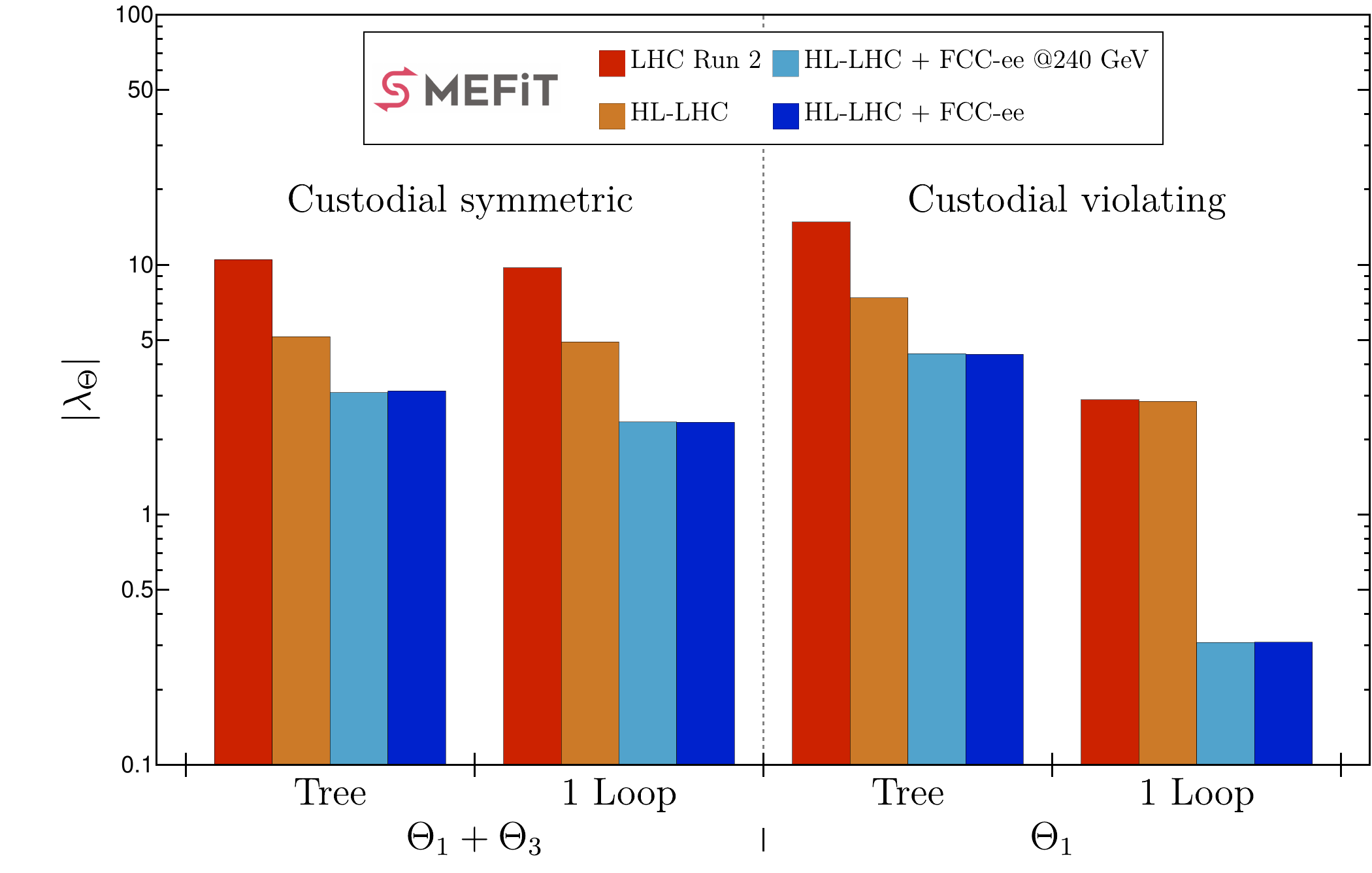}
    \caption{95$\%$ C.I. bounds on $|\lambda_{\Theta}|$ for the custodial quadruplets model, Eq.~(\ref{eq:quadruplet_lagrangian}), and the non-custodial model with only the $\Theta_{1}$ quadruplet, from a global quadratic SMEFT analysis based on the same datasets as Fig.~\ref{fig:bound_plot_cphi_hllhc_fcc} for both three-level and one-loop matching.
    }
    \label{fig:bounds_quadruplet}
\end{figure}

The 95\% C.I. on the Higgs-quadruplet coupling $|\lambda_{\Theta}|$ are reported in Fig.~\ref{fig:bounds_quadruplet} for the same dataset combinations as in Fig.~\ref{fig:bound_plot_cphi_hllhc_fcc}, both for the custodially-symmetric and custodially-violating variants, matched to the SMEFT to either tree- or one-loop level.
The tree-level matching results in both models showcase the (individual) sensitivity to the Higgs self-coupling at future colliders, with a factor of 2 improvement going from the LHC Run 2 to the HL-LHC and then a further $40\%$ reduction at FCC-ee.
This pattern of improvement follows from the individual bounds on $c_\varphi$ reported in Table~\ref{tab:bounds_all}, and are consistent with the result that for one-parameter fits no new constraints are provided by the $\sqrt{s}=365$ GeV run.

The one-loop matching affects both models differently.
In the non-custodial model, the bound tightens by an order of magnitude due to the sensitivity to $\mathcal{O}_{\varphi D}$, which more than compensates the factor $\sim 0.1$ between the matching coefficients for $c_{\varphi D}$ and $c_{\varphi}$.
One-loop matching improves the sensitivity to the custodial model by around $10\%$ and $25\%$ at the HL-LHC and FCC-ee respectively.
The improvement is caused by the tight bounds on $\mcO_{\varphi \Box}$, despite being generated with a coefficient much smaller than  $\mcO_\varphi$.
We note that this UV model correlates $c_{\varphi}$ and $c_{\varphi \Box}$ along a direction in the $(c_{\varphi},c_{\varphi \Box})$ plane poorly constrained at the FCC-ee unless one includes NNLO EW effects in EWPOs~\cite{Maura:2024zxz}.

\vspace{0.1cm}
\noindent
{\bf Summary and outlook.}
In this work we have revisited the sensitivity to the Higgs self-coupling in the global SMEFT analysis at the HL-LHC and the FCC-ee, in light of the most updated ESPPU26 projections.
Our main findings are that the FCC-ee with $\sqrt{s}\le 240$ GeV cannot improve significantly on the legacy HL-LHC constraints; that the addition of the $\sqrt{s}=365$ GeV run leads to a reduction of the bounds on $\kappa_3$ by a factor of 2; that individual and marginalized bounds are similar at the HL-LHC but can differ by up to a factor of 3 at the FCC-ee; and that quadratic EFT corrections are moderate but not negligible.
Our analysis demonstrates that the synergy of HL-LHC and FCC-ee data offers unique potential to pin down the Higgs self-coupling with $\sim$$15\%$ precision. 
These findings provide a timely input to the ESPPU26 by sharpening the constraints on the Higgs trilinear coupling achievable at the HL-LHC and FCC-ee. 

Our analysis could be extended by including higher order corrections to the predictions for the EWPOs and to the decays of the Higgs boson at the FCC-ee, as well as by exploring the impact of alternative flavour assumptions on the constraints obtained for $\kappa_3$. {Another area of future work concerns the addition of dimension-8 operators to our global SMEFT analysis, although their effect at the FCC-ee is expected to be small given the fact that the our linear and quadratic EFT projections are very similar there.}
Furthermore, it would be interesting to integrate in our analysis the constraints on di-Higgs production from high-energy lepton colliders such as LCF/CLIC and MuCol as well as those provided by the FCC-hh.

\vspace{0.05cm}
\paragraph{Acknowledgments.}
E. V. acknowledges helpful discussions with Gauthier Durieux and Fabio Maltoni. 
We are grateful to Freya Blekman, Valentina Cairo, and Pamela Ferrari for their input concerning di-Higgs measurements at the HL-LHC.

A. R. and E. V. are supported by the European Research Council (ERC) under the
European Union’s Horizon 2020 research and innovation programme (Grant agreement
No. 949451) and by a Royal Society University Research Fellowship through grant URF/R1/201553. The work of A. R. is also supported by the University of Padua under the
2023 STARS Grants@Unipd programme (Acronym and title of the project: HiggsPairs –
Precise Theoretical Predictions for Higgs pair production at the LHC). A. R. acknowledges
support from the COMETA COST Action CA22130 and  the
Mainz Institute for Theoretical Physics (MITP) of the Cluster of Excellence PRISMA+
(Project ID 390831469). L. M. acknowledges support from the European Union under the MSCA fellowship (Grant agreement N.
101149078) Advancing global SMEFT fits in the LHC precision era (EFT4ward). The
work of J. t. H. is supported by the UK Science and Technology Facility Council (STFC) consolidated grant ST/X000494/1. The work of J. R. is supported by the Dutch Research
Council (NWO) and by the Netherlands eScience Center.

\bibliography{Higgs_trilinear}

\begin{thebibliography}{51}
\expandafter\ifx\csname natexlab\endcsname\relax\def\natexlab#1{#1}\fi
\expandafter\ifx\csname bibnamefont\endcsname\relax
  \def\bibnamefont#1{#1}\fi
\expandafter\ifx\csname bibfnamefont\endcsname\relax
  \def\bibfnamefont#1{#1}\fi
\expandafter\ifx\csname citenamefont\endcsname\relax
  \def\citenamefont#1{#1}\fi
\expandafter\ifx\csname url\endcsname\relax
  \def\url#1{\texttt{#1}}\fi
\expandafter\ifx\csname urlprefix\endcsname\relax\def\urlprefix{URL }\fi
\providecommand{\bibinfo}[2]{#2}
\providecommand{\eprint}[2][]{\url{#2}}

\bibitem[{\citenamefont{Cepeda et~al.}(2019)}]{Cepeda:2019klc}
\bibinfo{author}{\bibfnamefont{M.}~\bibnamefont{Cepeda}} \bibnamefont{et~al.},
  \bibinfo{journal}{CERN Yellow Rep. Monogr.} \textbf{\bibinfo{volume}{7}},
  \bibinfo{pages}{221} (\bibinfo{year}{2019}), \eprint{1902.00134}.

\bibitem[{\citenamefont{Atlas}(2025)}]{Atlas:2025kye}
\bibinfo{author}{\bibnamefont{Atlas}} (\bibinfo{collaboration}{CMS})
  (\bibinfo{year}{2025}), \eprint{2504.00672}.

\bibitem[{\citenamefont{Abada et~al.}(2019{\natexlab{a}})}]{FCC:2018evy}
\bibinfo{author}{\bibfnamefont{A.}~\bibnamefont{Abada}} \bibnamefont{et~al.}
  (\bibinfo{collaboration}{FCC}), \bibinfo{journal}{Eur. Phys. J. ST}
  \textbf{\bibinfo{volume}{228}}, \bibinfo{pages}{261}
  (\bibinfo{year}{2019}{\natexlab{a}}).

\bibitem[{\citenamefont{Benedikt et~al.}(2025)\citenamefont{Benedikt, Bartmann,
  Burnet, Carli, Chance, Craievich, Giovannozzi, Grojean, Gutleber, Hanke
  et~al.}}]{Benedikt:2928193}
\bibinfo{author}{\bibfnamefont{M.}~\bibnamefont{Benedikt}},
  \bibinfo{author}{\bibfnamefont{W.}~\bibnamefont{Bartmann}},
  \bibinfo{author}{\bibfnamefont{J.-P.} \bibnamefont{Burnet}},
  \bibinfo{author}{\bibfnamefont{C.}~\bibnamefont{Carli}},
  \bibinfo{author}{\bibfnamefont{A.}~\bibnamefont{Chance}},
  \bibinfo{author}{\bibfnamefont{P.}~\bibnamefont{Craievich}},
  \bibinfo{author}{\bibfnamefont{M.}~\bibnamefont{Giovannozzi}},
  \bibinfo{author}{\bibfnamefont{C.}~\bibnamefont{Grojean}},
  \bibinfo{author}{\bibfnamefont{J.}~\bibnamefont{Gutleber}},
  \bibinfo{author}{\bibfnamefont{K.}~\bibnamefont{Hanke}},
  \bibnamefont{et~al.}, \bibinfo{journal}{CERN-FCC-PHYS-2025-0002}
  (\bibinfo{year}{2025}), \eprint{https://cds.cern.ch/record/2928193}.

\bibitem[{\citenamefont{Abada et~al.}(2019{\natexlab{b}})}]{FCC:2018vvp}
\bibinfo{author}{\bibfnamefont{A.}~\bibnamefont{Abada}} \bibnamefont{et~al.}
  (\bibinfo{collaboration}{FCC}), \bibinfo{journal}{Eur. Phys. J. ST}
  \textbf{\bibinfo{volume}{228}}, \bibinfo{pages}{755}
  (\bibinfo{year}{2019}{\natexlab{b}}).

\bibitem[{\citenamefont{Baer et~al.}(2013)}]{ILC:2013jhg}
\bibinfo{author}{\bibfnamefont{H.}~\bibnamefont{Baer}} \bibnamefont{et~al.}
  (\bibinfo{collaboration}{ILC}) (\bibinfo{year}{2013}), \eprint{1306.6352}.

\bibitem[{\citenamefont{Abramowicz
  et~al.}(2025)}]{LinearColliderVision:2025hlt}
\bibinfo{author}{\bibfnamefont{H.}~\bibnamefont{Abramowicz}}
  \bibnamefont{et~al.} (\bibinfo{collaboration}{Linear Collider Vision})
  (\bibinfo{year}{2025}), \eprint{2503.19983}.

\bibitem[{\citenamefont{Cheng et~al.}(2022)}]{CEPCPhysicsStudyGroup:2022uwl}
\bibinfo{author}{\bibfnamefont{H.}~\bibnamefont{Cheng}} \bibnamefont{et~al.}
  (\bibinfo{collaboration}{CEPC Physics Study Group}), in
  \emph{\bibinfo{booktitle}{{Snowmass 2021}}} (\bibinfo{year}{2022}),
  \eprint{2205.08553}.

\bibitem[{\citenamefont{Linssen et~al.}(2012)\citenamefont{Linssen, Miyamoto,
  Stanitzki, and Weerts}}]{Linssen:2012hp}
\bibinfo{author}{\bibfnamefont{L.}~\bibnamefont{Linssen}},
  \bibinfo{author}{\bibfnamefont{A.}~\bibnamefont{Miyamoto}},
  \bibinfo{author}{\bibfnamefont{M.}~\bibnamefont{Stanitzki}},
  \bibnamefont{and} \bibinfo{author}{\bibfnamefont{H.}~\bibnamefont{Weerts}}
  (\bibinfo{collaboration}{CLIC}) (\bibinfo{year}{2012}), \eprint{1202.5940}.

\bibitem[{Aic(2012)}]{Aicheler:2012bya}
 (\bibinfo{year}{2012}), \eprint{10.5170/CERN-2012-007}.

\bibitem[{\citenamefont{Accettura et~al.}(2023)}]{Accettura:2023ked}
\bibinfo{author}{\bibfnamefont{C.}~\bibnamefont{Accettura}}
  \bibnamefont{et~al.}, \bibinfo{journal}{Eur. Phys. J. C}
  \textbf{\bibinfo{volume}{83}}, \bibinfo{pages}{864} (\bibinfo{year}{2023}),
  \bibinfo{note}{[Erratum: Eur.Phys.J.C 84, 36 (2024)]}, \eprint{2303.08533}.

\bibitem[{\citenamefont{Accettura
  et~al.}(2024)}]{InternationalMuonCollider:2024jyv}
\bibinfo{author}{\bibfnamefont{C.}~\bibnamefont{Accettura}}
  \bibnamefont{et~al.} (\bibinfo{collaboration}{International Muon Collider}),
  \bibinfo{journal}{CERN Yellow Rep. Monogr.}
  \textbf{\bibinfo{volume}{2/2024}}, \bibinfo{pages}{176}
  (\bibinfo{year}{2024}), \eprint{2407.12450}.

\bibitem[{\citenamefont{Hiller et~al.}(2024)\citenamefont{Hiller, H\"ohne,
  Litim, and Steudtner}}]{Hiller:2024zjp}
\bibinfo{author}{\bibfnamefont{G.}~\bibnamefont{Hiller}},
  \bibinfo{author}{\bibfnamefont{T.}~\bibnamefont{H\"ohne}},
  \bibinfo{author}{\bibfnamefont{D.~F.} \bibnamefont{Litim}}, \bibnamefont{and}
  \bibinfo{author}{\bibfnamefont{T.}~\bibnamefont{Steudtner}},
  \bibinfo{journal}{Phys. Rev. D} \textbf{\bibinfo{volume}{110}},
  \bibinfo{pages}{115017} (\bibinfo{year}{2024}), \eprint{2401.08811}.

\bibitem[{\citenamefont{Andreassen et~al.}(2018)\citenamefont{Andreassen,
  Frost, and Schwartz}}]{Andreassen:2017rzq}
\bibinfo{author}{\bibfnamefont{A.}~\bibnamefont{Andreassen}},
  \bibinfo{author}{\bibfnamefont{W.}~\bibnamefont{Frost}}, \bibnamefont{and}
  \bibinfo{author}{\bibfnamefont{M.~D.} \bibnamefont{Schwartz}},
  \bibinfo{journal}{Phys. Rev. D} \textbf{\bibinfo{volume}{97}},
  \bibinfo{pages}{056006} (\bibinfo{year}{2018}), \eprint{1707.08124}.

\bibitem[{\citenamefont{Zhang et~al.}(2023)\citenamefont{Zhang, Li, Liu,
  Ramsey-Musolf, Zeng, and Arunasalam}}]{Zhang:2023jvh}
\bibinfo{author}{\bibfnamefont{W.}~\bibnamefont{Zhang}},
  \bibinfo{author}{\bibfnamefont{H.-L.} \bibnamefont{Li}},
  \bibinfo{author}{\bibfnamefont{K.}~\bibnamefont{Liu}},
  \bibinfo{author}{\bibfnamefont{M.~J.} \bibnamefont{Ramsey-Musolf}},
  \bibinfo{author}{\bibfnamefont{Y.}~\bibnamefont{Zeng}}, \bibnamefont{and}
  \bibinfo{author}{\bibfnamefont{S.}~\bibnamefont{Arunasalam}},
  \bibinfo{journal}{JHEP} \textbf{\bibinfo{volume}{12}}, \bibinfo{pages}{018}
  (\bibinfo{year}{2023}), \eprint{2303.03612}.

\bibitem[{\citenamefont{Huang et~al.}(2016)\citenamefont{Huang, Long, and
  Wang}}]{Huang:2016cjm}
\bibinfo{author}{\bibfnamefont{P.}~\bibnamefont{Huang}},
  \bibinfo{author}{\bibfnamefont{A.~J.} \bibnamefont{Long}}, \bibnamefont{and}
  \bibinfo{author}{\bibfnamefont{L.-T.} \bibnamefont{Wang}},
  \bibinfo{journal}{Phys. Rev. D} \textbf{\bibinfo{volume}{94}},
  \bibinfo{pages}{075008} (\bibinfo{year}{2016}), \eprint{1608.06619}.

\bibitem[{\citenamefont{Bednyakov et~al.}(2015)\citenamefont{Bednyakov, Kniehl,
  Pikelner, and Veretin}}]{Bednyakov:2015sca}
\bibinfo{author}{\bibfnamefont{A.~V.} \bibnamefont{Bednyakov}},
  \bibinfo{author}{\bibfnamefont{B.~A.} \bibnamefont{Kniehl}},
  \bibinfo{author}{\bibfnamefont{A.~F.} \bibnamefont{Pikelner}},
  \bibnamefont{and} \bibinfo{author}{\bibfnamefont{O.~L.}
  \bibnamefont{Veretin}}, \bibinfo{journal}{Phys. Rev. Lett.}
  \textbf{\bibinfo{volume}{115}}, \bibinfo{pages}{201802}
  (\bibinfo{year}{2015}), \eprint{1507.08833}.

\bibitem[{\citenamefont{Isidori et~al.}(2024)\citenamefont{Isidori, Wilsch, and
  Wyler}}]{Isidori:2023pyp}
\bibinfo{author}{\bibfnamefont{G.}~\bibnamefont{Isidori}},
  \bibinfo{author}{\bibfnamefont{F.}~\bibnamefont{Wilsch}}, \bibnamefont{and}
  \bibinfo{author}{\bibfnamefont{D.}~\bibnamefont{Wyler}},
  \bibinfo{journal}{Rev. Mod. Phys.} \textbf{\bibinfo{volume}{96}},
  \bibinfo{pages}{015006} (\bibinfo{year}{2024}), \eprint{2303.16922}.

\bibitem[{\citenamefont{Alasfar et~al.}(2024)}]{Alasfar:2023xpc}
\bibinfo{author}{\bibfnamefont{L.}~\bibnamefont{Alasfar}} \bibnamefont{et~al.},
  \bibinfo{journal}{SciPost Phys. Comm. Rep.} \textbf{\bibinfo{volume}{2024}},
  \bibinfo{pages}{2} (\bibinfo{year}{2024}), \eprint{2304.01968}.

\bibitem[{\citenamefont{Aad et~al.}(2024)}]{ATLAS:2024ish}
\bibinfo{author}{\bibfnamefont{G.}~\bibnamefont{Aad}} \bibnamefont{et~al.}
  (\bibinfo{collaboration}{ATLAS}), \bibinfo{journal}{Phys. Rev. Lett.}
  \textbf{\bibinfo{volume}{133}}, \bibinfo{pages}{101801}
  (\bibinfo{year}{2024}), \eprint{2406.09971}.

\bibitem[{\citenamefont{Hayrapetyan et~al.}(2025)}]{CMS:2024awa}
\bibinfo{author}{\bibfnamefont{A.}~\bibnamefont{Hayrapetyan}}
  \bibnamefont{et~al.} (\bibinfo{collaboration}{CMS}), \bibinfo{journal}{Phys.
  Lett. B} \textbf{\bibinfo{volume}{861}}, \bibinfo{pages}{139210}
  (\bibinfo{year}{2025}), \eprint{2407.13554}.

\bibitem[{\citenamefont{Gorbahn and Haisch}(2016)}]{Gorbahn:2016uoy}
\bibinfo{author}{\bibfnamefont{M.}~\bibnamefont{Gorbahn}} \bibnamefont{and}
  \bibinfo{author}{\bibfnamefont{U.}~\bibnamefont{Haisch}},
  \bibinfo{journal}{JHEP} \textbf{\bibinfo{volume}{10}}, \bibinfo{pages}{094}
  (\bibinfo{year}{2016}), \eprint{1607.03773}.

\bibitem[{\citenamefont{Maltoni et~al.}(2017)\citenamefont{Maltoni, Pagani,
  Shivaji, and Zhao}}]{Maltoni:2017ims}
\bibinfo{author}{\bibfnamefont{F.}~\bibnamefont{Maltoni}},
  \bibinfo{author}{\bibfnamefont{D.}~\bibnamefont{Pagani}},
  \bibinfo{author}{\bibfnamefont{A.}~\bibnamefont{Shivaji}}, \bibnamefont{and}
  \bibinfo{author}{\bibfnamefont{X.}~\bibnamefont{Zhao}},
  \bibinfo{journal}{Eur. Phys. J. C} \textbf{\bibinfo{volume}{77}},
  \bibinfo{pages}{887} (\bibinfo{year}{2017}), \eprint{1709.08649}.

\bibitem[{\citenamefont{Degrassi et~al.}(2016)\citenamefont{Degrassi, Giardino,
  Maltoni, and Pagani}}]{Degrassi:2016wml}
\bibinfo{author}{\bibfnamefont{G.}~\bibnamefont{Degrassi}},
  \bibinfo{author}{\bibfnamefont{P.~P.} \bibnamefont{Giardino}},
  \bibinfo{author}{\bibfnamefont{F.}~\bibnamefont{Maltoni}}, \bibnamefont{and}
  \bibinfo{author}{\bibfnamefont{D.}~\bibnamefont{Pagani}},
  \bibinfo{journal}{JHEP} \textbf{\bibinfo{volume}{12}}, \bibinfo{pages}{080}
  (\bibinfo{year}{2016}), \eprint{1607.04251}.

\bibitem[{\citenamefont{Bizon et~al.}(2017)\citenamefont{Bizon, Gorbahn,
  Haisch, and Zanderighi}}]{Bizon:2016wgr}
\bibinfo{author}{\bibfnamefont{W.}~\bibnamefont{Bizon}},
  \bibinfo{author}{\bibfnamefont{M.}~\bibnamefont{Gorbahn}},
  \bibinfo{author}{\bibfnamefont{U.}~\bibnamefont{Haisch}}, \bibnamefont{and}
  \bibinfo{author}{\bibfnamefont{G.}~\bibnamefont{Zanderighi}},
  \bibinfo{journal}{JHEP} \textbf{\bibinfo{volume}{07}}, \bibinfo{pages}{083}
  (\bibinfo{year}{2017}), \eprint{1610.05771}.

\bibitem[{\citenamefont{Anastopoulos et~al.}(2025)}]{Anastopoulos:2025jyh}
\bibinfo{author}{\bibfnamefont{C.}~\bibnamefont{Anastopoulos}}
  \bibnamefont{et~al.} (\bibinfo{year}{2025}), \eprint{2504.00541}.

\bibitem[{\citenamefont{McCullough}(2014)}]{McCullough:2013rea}
\bibinfo{author}{\bibfnamefont{M.}~\bibnamefont{McCullough}},
  \bibinfo{journal}{Phys. Rev. D} \textbf{\bibinfo{volume}{90}},
  \bibinfo{pages}{015001} (\bibinfo{year}{2014}), \bibinfo{note}{[Erratum:
  Phys.Rev.D 92, 039903 (2015)]}, \eprint{1312.3322}.

\bibitem[{\citenamefont{Maltoni et~al.}(2018)\citenamefont{Maltoni, Pagani, and
  Zhao}}]{Maltoni:2018ttu}
\bibinfo{author}{\bibfnamefont{F.}~\bibnamefont{Maltoni}},
  \bibinfo{author}{\bibfnamefont{D.}~\bibnamefont{Pagani}}, \bibnamefont{and}
  \bibinfo{author}{\bibfnamefont{X.}~\bibnamefont{Zhao}},
  \bibinfo{journal}{JHEP} \textbf{\bibinfo{volume}{07}}, \bibinfo{pages}{087}
  (\bibinfo{year}{2018}), \eprint{1802.07616}.

\bibitem[{\citenamefont{Asteriadis et~al.}(2024)\citenamefont{Asteriadis,
  Dawson, Giardino, and Szafron}}]{Asteriadis:2024xuk}
\bibinfo{author}{\bibfnamefont{K.}~\bibnamefont{Asteriadis}},
  \bibinfo{author}{\bibfnamefont{S.}~\bibnamefont{Dawson}},
  \bibinfo{author}{\bibfnamefont{P.~P.} \bibnamefont{Giardino}},
  \bibnamefont{and} \bibinfo{author}{\bibfnamefont{R.}~\bibnamefont{Szafron}},
  \bibinfo{journal}{Phys. Rev. Lett.} \textbf{\bibinfo{volume}{133}},
  \bibinfo{pages}{231801} (\bibinfo{year}{2024}), \eprint{2406.03557}.

\bibitem[{\citenamefont{Asteriadis et~al.}(2025)\citenamefont{Asteriadis,
  Dawson, Giardino, and Szafron}}]{Asteriadis:2024xts}
\bibinfo{author}{\bibfnamefont{K.}~\bibnamefont{Asteriadis}},
  \bibinfo{author}{\bibfnamefont{S.}~\bibnamefont{Dawson}},
  \bibinfo{author}{\bibfnamefont{P.~P.} \bibnamefont{Giardino}},
  \bibnamefont{and} \bibinfo{author}{\bibfnamefont{R.}~\bibnamefont{Szafron}},
  \bibinfo{journal}{JHEP} \textbf{\bibinfo{volume}{02}}, \bibinfo{pages}{162}
  (\bibinfo{year}{2025}), \eprint{2409.11466}.

\bibitem[{\citenamefont{Maura et~al.}(2025)\citenamefont{Maura, Stefanek, and
  You}}]{Maura:2025rcv}
\bibinfo{author}{\bibfnamefont{V.}~\bibnamefont{Maura}},
  \bibinfo{author}{\bibfnamefont{B.~A.} \bibnamefont{Stefanek}},
  \bibnamefont{and} \bibinfo{author}{\bibfnamefont{T.}~\bibnamefont{You}}
  (\bibinfo{year}{2025}), \eprint{2503.13719}.

\bibitem[{\citenamefont{de~Blas et~al.}(2020)}]{deBlas:2019rxi}
\bibinfo{author}{\bibfnamefont{J.}~\bibnamefont{de~Blas}} \bibnamefont{et~al.},
  \bibinfo{journal}{JHEP} \textbf{\bibinfo{volume}{01}}, \bibinfo{pages}{139}
  (\bibinfo{year}{2020}), \eprint{1905.03764}.

\bibitem[{\citenamefont{Collaboration}(2023)}]{2780507}
\bibinfo{author}{\bibfnamefont{F.}~\bibnamefont{Collaboration}}
  (\bibinfo{collaboration}{FCC}), \bibinfo{journal}{CERN/FC/6734/RA}
  (\bibinfo{year}{2023}), \eprint{https://cds.cern.ch/record/2887250}.

\bibitem[{\citenamefont{Celada et~al.}(2024)\citenamefont{Celada, Giani, ter
  Hoeve, Mantani, Rojo, Rossia, Thomas, and Vryonidou}}]{Celada:2024mcf}
\bibinfo{author}{\bibfnamefont{E.}~\bibnamefont{Celada}},
  \bibinfo{author}{\bibfnamefont{T.}~\bibnamefont{Giani}},
  \bibinfo{author}{\bibfnamefont{J.}~\bibnamefont{ter Hoeve}},
  \bibinfo{author}{\bibfnamefont{L.}~\bibnamefont{Mantani}},
  \bibinfo{author}{\bibfnamefont{J.}~\bibnamefont{Rojo}},
  \bibinfo{author}{\bibfnamefont{A.~N.} \bibnamefont{Rossia}},
  \bibinfo{author}{\bibfnamefont{M.~O.~A.} \bibnamefont{Thomas}},
  \bibnamefont{and}
  \bibinfo{author}{\bibfnamefont{E.}~\bibnamefont{Vryonidou}},
  \bibinfo{journal}{JHEP} \textbf{\bibinfo{volume}{09}}, \bibinfo{pages}{091}
  (\bibinfo{year}{2024}), \eprint{2404.12809}.

\bibitem[{\citenamefont{ter Hoeve et~al.}(2025)\citenamefont{ter Hoeve,
  Mantani, Rossia, Rojo, and Vryonidou}}]{terHoeve:2025gey}
\bibinfo{author}{\bibfnamefont{J.}~\bibnamefont{ter Hoeve}},
  \bibinfo{author}{\bibfnamefont{L.}~\bibnamefont{Mantani}},
  \bibinfo{author}{\bibfnamefont{A.~N.} \bibnamefont{Rossia}},
  \bibinfo{author}{\bibfnamefont{J.}~\bibnamefont{Rojo}}, \bibnamefont{and}
  \bibinfo{author}{\bibfnamefont{E.}~\bibnamefont{Vryonidou}}
  (\bibinfo{year}{2025}), \eprint{2502.20453}.

\bibitem[{\citenamefont{Durieux et~al.}(2022)\citenamefont{Durieux, McCullough,
  and Salvioni}}]{Durieux:2022hbu}
\bibinfo{author}{\bibfnamefont{G.}~\bibnamefont{Durieux}},
  \bibinfo{author}{\bibfnamefont{M.}~\bibnamefont{McCullough}},
  \bibnamefont{and} \bibinfo{author}{\bibfnamefont{E.}~\bibnamefont{Salvioni}},
  \bibinfo{journal}{JHEP} \textbf{\bibinfo{volume}{12}}, \bibinfo{pages}{148}
  (\bibinfo{year}{2022}), \bibinfo{note}{[Erratum: JHEP 02, 165 (2023)]},
  \eprint{2209.00666}.

\bibitem[{\citenamefont{Hartland et~al.}(2019)\citenamefont{Hartland, Maltoni,
  Nocera, Rojo, Slade, Vryonidou, and Zhang}}]{Hartland:2019bjb}
\bibinfo{author}{\bibfnamefont{N.~P.} \bibnamefont{Hartland}},
  \bibinfo{author}{\bibfnamefont{F.}~\bibnamefont{Maltoni}},
  \bibinfo{author}{\bibfnamefont{E.~R.} \bibnamefont{Nocera}},
  \bibinfo{author}{\bibfnamefont{J.}~\bibnamefont{Rojo}},
  \bibinfo{author}{\bibfnamefont{E.}~\bibnamefont{Slade}},
  \bibinfo{author}{\bibfnamefont{E.}~\bibnamefont{Vryonidou}},
  \bibnamefont{and} \bibinfo{author}{\bibfnamefont{C.}~\bibnamefont{Zhang}},
  \bibinfo{journal}{JHEP} \textbf{\bibinfo{volume}{04}}, \bibinfo{pages}{100}
  (\bibinfo{year}{2019}), \eprint{1901.05965}.

\bibitem[{\citenamefont{Ethier et~al.}(2021{\natexlab{a}})\citenamefont{Ethier,
  Gomez-Ambrosio, Magni, and Rojo}}]{Ethier:2021ydt}
\bibinfo{author}{\bibfnamefont{J.~J.} \bibnamefont{Ethier}},
  \bibinfo{author}{\bibfnamefont{R.}~\bibnamefont{Gomez-Ambrosio}},
  \bibinfo{author}{\bibfnamefont{G.}~\bibnamefont{Magni}}, \bibnamefont{and}
  \bibinfo{author}{\bibfnamefont{J.}~\bibnamefont{Rojo}},
  \bibinfo{journal}{Eur. Phys. J. C} \textbf{\bibinfo{volume}{81}},
  \bibinfo{pages}{560} (\bibinfo{year}{2021}{\natexlab{a}}),
  \eprint{2101.03180}.

\bibitem[{\citenamefont{Giani et~al.}(2023)\citenamefont{Giani, Magni, and
  Rojo}}]{Giani:2023gfq}
\bibinfo{author}{\bibfnamefont{T.}~\bibnamefont{Giani}},
  \bibinfo{author}{\bibfnamefont{G.}~\bibnamefont{Magni}}, \bibnamefont{and}
  \bibinfo{author}{\bibfnamefont{J.}~\bibnamefont{Rojo}},
  \bibinfo{journal}{Eur. Phys. J. C} \textbf{\bibinfo{volume}{83}},
  \bibinfo{pages}{393} (\bibinfo{year}{2023}), \eprint{2302.06660}.

\bibitem[{\citenamefont{Ethier et~al.}(2021{\natexlab{b}})\citenamefont{Ethier,
  Magni, Maltoni, Mantani, Nocera, Rojo, Slade, Vryonidou, and
  Zhang}}]{Ethier:2021bye}
\bibinfo{author}{\bibfnamefont{J.~J.} \bibnamefont{Ethier}},
  \bibinfo{author}{\bibfnamefont{G.}~\bibnamefont{Magni}},
  \bibinfo{author}{\bibfnamefont{F.}~\bibnamefont{Maltoni}},
  \bibinfo{author}{\bibfnamefont{L.}~\bibnamefont{Mantani}},
  \bibinfo{author}{\bibfnamefont{E.~R.} \bibnamefont{Nocera}},
  \bibinfo{author}{\bibfnamefont{J.}~\bibnamefont{Rojo}},
  \bibinfo{author}{\bibfnamefont{E.}~\bibnamefont{Slade}},
  \bibinfo{author}{\bibfnamefont{E.}~\bibnamefont{Vryonidou}},
  \bibnamefont{and} \bibinfo{author}{\bibfnamefont{C.}~\bibnamefont{Zhang}}
  (\bibinfo{collaboration}{SMEFiT}), \bibinfo{journal}{JHEP}
  \textbf{\bibinfo{volume}{11}}, \bibinfo{pages}{089}
  (\bibinfo{year}{2021}{\natexlab{b}}), \eprint{2105.00006}.

\bibitem[{\citenamefont{ter Hoeve et~al.}(2024)\citenamefont{ter Hoeve, Magni,
  Rojo, Rossia, and Vryonidou}}]{terHoeve:2023pvs}
\bibinfo{author}{\bibfnamefont{J.}~\bibnamefont{ter Hoeve}},
  \bibinfo{author}{\bibfnamefont{G.}~\bibnamefont{Magni}},
  \bibinfo{author}{\bibfnamefont{J.}~\bibnamefont{Rojo}},
  \bibinfo{author}{\bibfnamefont{A.~N.} \bibnamefont{Rossia}},
  \bibnamefont{and}
  \bibinfo{author}{\bibfnamefont{E.}~\bibnamefont{Vryonidou}},
  \bibinfo{journal}{JHEP} \textbf{\bibinfo{volume}{01}}, \bibinfo{pages}{179}
  (\bibinfo{year}{2024}), \eprint{2309.04523}.

\bibitem[{\citenamefont{ATLAS}(2025{\natexlab{a}})}]{ATL-PHYS-PUB-2025-006}
\bibinfo{author}{\bibnamefont{ATLAS}} (\bibinfo{year}{2025}{\natexlab{a}}),
  \eprint{ATL-PHYS-PUB-2025-006}.

\bibitem[{\citenamefont{ATLAS}(2024)}]{ATL-PHYS-PUB-2024-016}
\bibinfo{author}{\bibnamefont{ATLAS}} (\bibinfo{year}{2024}),
  \eprint{ATL-PHYS-PUB-2024-016}.

\bibitem[{\citenamefont{ATLAS}(2025{\natexlab{b}})}]{ATL-PHYS-PUB-2025-001}
\bibinfo{author}{\bibnamefont{ATLAS}} (\bibinfo{year}{2025}{\natexlab{b}}),
  \eprint{ATL-PHYS-PUB-2025-001}.

\bibitem[{\citenamefont{Collaboration}(2025)}]{Collaboration:2928096}
\bibinfo{author}{\bibfnamefont{C.}~\bibnamefont{Collaboration}}
  (\bibinfo{collaboration}{CMS}) (\bibinfo{year}{2025}),
  \bibinfo{note}{cMS-NOTE-2025-006},
  \eprint{https://cds.cern.ch/record/2928096/}.

\bibitem[{\citenamefont{Di~Luzio et~al.}(2017)\citenamefont{Di~Luzio, Gr\"ober,
  and Spannowsky}}]{DiLuzio:2017tfn}
\bibinfo{author}{\bibfnamefont{L.}~\bibnamefont{Di~Luzio}},
  \bibinfo{author}{\bibfnamefont{R.}~\bibnamefont{Gr\"ober}}, \bibnamefont{and}
  \bibinfo{author}{\bibfnamefont{M.}~\bibnamefont{Spannowsky}},
  \bibinfo{journal}{Eur. Phys. J. C} \textbf{\bibinfo{volume}{77}},
  \bibinfo{pages}{788} (\bibinfo{year}{2017}), \eprint{1704.02311}.

\bibitem[{\citenamefont{Falkowski and Rattazzi}(2019)}]{Falkowski:2019tft}
\bibinfo{author}{\bibfnamefont{A.}~\bibnamefont{Falkowski}} \bibnamefont{and}
  \bibinfo{author}{\bibfnamefont{R.}~\bibnamefont{Rattazzi}},
  \bibinfo{journal}{JHEP} \textbf{\bibinfo{volume}{10}}, \bibinfo{pages}{255}
  (\bibinfo{year}{2019}), \eprint{1902.05936}.

\bibitem[{\citenamefont{de~Blas et~al.}(2018)\citenamefont{de~Blas, Criado,
  Perez-Victoria, and Santiago}}]{deBlas:2017xtg}
\bibinfo{author}{\bibfnamefont{J.}~\bibnamefont{de~Blas}},
  \bibinfo{author}{\bibfnamefont{J.~C.} \bibnamefont{Criado}},
  \bibinfo{author}{\bibfnamefont{M.}~\bibnamefont{Perez-Victoria}},
  \bibnamefont{and} \bibinfo{author}{\bibfnamefont{J.}~\bibnamefont{Santiago}},
  \bibinfo{journal}{JHEP} \textbf{\bibinfo{volume}{03}}, \bibinfo{pages}{109}
  (\bibinfo{year}{2018}), \eprint{1711.10391}.

\bibitem[{\citenamefont{Maura et~al.}(2024)\citenamefont{Maura, Stefanek, and
  You}}]{Maura:2024zxz}
\bibinfo{author}{\bibfnamefont{V.}~\bibnamefont{Maura}},
  \bibinfo{author}{\bibfnamefont{B.~A.} \bibnamefont{Stefanek}},
  \bibnamefont{and} \bibinfo{author}{\bibfnamefont{T.}~\bibnamefont{You}}
  (\bibinfo{year}{2024}), \eprint{2412.14241}.

\bibitem[{\citenamefont{Alison et~al.}(2020)}]{DiMicco:2019ngk}
\bibinfo{author}{\bibfnamefont{J.}~\bibnamefont{Alison}} \bibnamefont{et~al.},
  \bibinfo{journal}{Rev. Phys.} \textbf{\bibinfo{volume}{5}},
  \bibinfo{pages}{100045} (\bibinfo{year}{2020}), \eprint{1910.00012}.

\bibitem[{\citenamefont{Di~Vita et~al.}(2018)\citenamefont{Di~Vita, Durieux,
  Grojean, Gu, Liu, Panico, Riembau, and Vantalon}}]{DiVita:2017vrr}
\bibinfo{author}{\bibfnamefont{S.}~\bibnamefont{Di~Vita}},
  \bibinfo{author}{\bibfnamefont{G.}~\bibnamefont{Durieux}},
  \bibinfo{author}{\bibfnamefont{C.}~\bibnamefont{Grojean}},
  \bibinfo{author}{\bibfnamefont{J.}~\bibnamefont{Gu}},
  \bibinfo{author}{\bibfnamefont{Z.}~\bibnamefont{Liu}},
  \bibinfo{author}{\bibfnamefont{G.}~\bibnamefont{Panico}},
  \bibinfo{author}{\bibfnamefont{M.}~\bibnamefont{Riembau}}, \bibnamefont{and}
  \bibinfo{author}{\bibfnamefont{T.}~\bibnamefont{Vantalon}},
  \bibinfo{journal}{JHEP} \textbf{\bibinfo{volume}{02}}, \bibinfo{pages}{178}
  (\bibinfo{year}{2018}), \eprint{1711.03978}.

\end{thebibliography}

\onecolumngrid
\appendix

\begin{center}
\LARGE
{\bf Supporting Information}
\end{center}

\section{The HL-LHC likelihood for Higgs pair production}
\label{app:likelihood}

In this appendix we describe our implementation of the ATLAS+CMS combined likelihood for Higgs pair production at the HL-LHC for $\mathcal{L}=3$ ab$^{-1}$, based on the latest projections prepared in the context of the ESPPU26~\cite{Atlas:2025kye}.
For the systematic uncertainties, we work in the so-called scenario S3, described in~\cite{Collaboration:2928096,ATL-PHYS-PUB-2025-006}, which extends the previously used scenario S2~\cite{Cepeda:2019klc} by considering the 5\% improvement in $b$-jet tagging and hadronic $\tau$ reconstruction efficiencies which have already been achieved at Run 3. 
As compared to the previous (2019) ESPPU projections~\cite{Cepeda:2019klc}, the main improvements in the ESPPU26 ones are those  associated to the higher efficiency of the flavoured-jet tagging algorithms used for $b$-jet reconstruction. 

Our likelihood for Higgs pair production at the HL-LHC is constructed by combining the projected measurement of the di-Higgs signal strength in ATLAS in S3, $\delta\mu_{hh}=_{-25\%}^{+28\%}$ at the 68\% C.L.~\cite{ATL-PHYS-PUB-2025-006}, with the differential information in the $b\bar{b}\gamma\gamma$~\cite{ATL-PHYS-PUB-2025-001} and the $b\bar{b}\tau^+\tau^-$~\cite{ATL-PHYS-PUB-2024-016} final states also from ATLAS.
While CMS does not provide an explicit projection for $\mu_{hh}$ in~\cite{Collaboration:2928096}, their expected significance for $hh$ production is 4.5$\sigma$ in S3, which is the same as ATLAS~\cite{ATL-PHYS-PUB-2025-006}, and hence it is justified to assume the same value of $\delta\mu_{hh}$ for CMS as for ATLAS.
Concerning the differential measurements, as demonstrated in~\cite{ATL-PHYS-PUB-2025-006,Collaboration:2928096}, the overall sensitivity to Higgs pair production at the HL-LHC is dominated by the contribution from the two final states considered, $b\bar{b}\gamma\gamma$ and $b\bar{b}\tau^+\tau^-$.  
As well known, the inclusion of differential information is necessary to eliminate a second spurious solution in the quadratic EFT fit~\cite{DiMicco:2019ngk,DiVita:2017vrr}.
The ATLAS $b\bar{b}\gamma\gamma$~\cite{ATL-PHYS-PUB-2025-001} and $b\bar{b}\tau^+\tau^-$~\cite{ATL-PHYS-PUB-2024-016} analysis are provided in two bins of invariant mass, one with $m_{hh}> 350$ GeV, which dominates the SM signal strength, and the other one with $m_{hh}<350$ GeV, which targets BSM applications. 

\begin{figure}[h]
    \centering
    \includegraphics[width=0.750\textwidth]{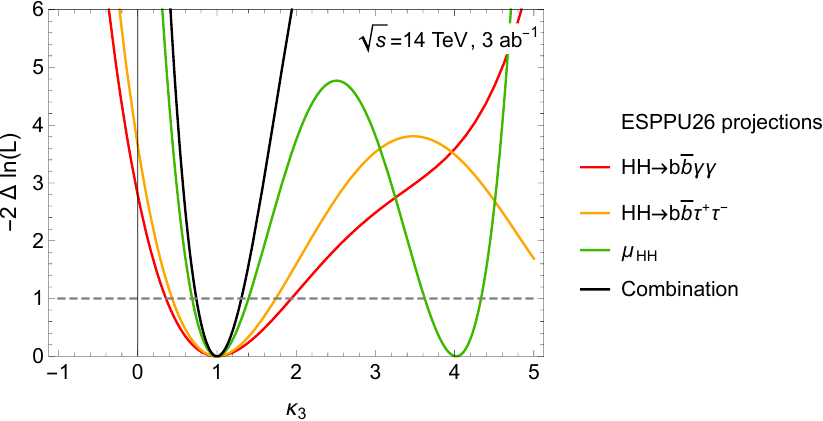}
    \caption{
    The log-likelihood $\Delta\chi^2(\kappa_3) = -2\Delta\ln L(\kappa_3)$ as a function of the  Higgs self-coupling modifier $\kappa_3$ for Higgs pair-production at the HL-LHC corresponding to  $\mathcal{L}=3$ ab$^{-1}$.
    Here all SMEFT coefficients other than $c_\varphi$ are set to zero.
    We display the likelihoods for the differential analyses of $hh\to b\bar{b}\gamma\gamma$ and $hh\to b\bar{b}\tau^+\tau^-$, for the inclusive signal strength $\mu_{hh}$, and for their combination.
    The horizontal dashed line indicates the $\Delta\chi^2=1$ boundary.
      }
\label{fig:likelihood_dihiggs_hllhc}
\end{figure}

Fig.~\ref{fig:likelihood_dihiggs_hllhc} displays the resulting log-likelihood $\chi^2(\kappa_\lambda)=-2\ln L(\kappa_\lambda)$ as a function of the  Higgs self-coupling modifier $\kappa_\lambda$ for Higgs pair-production at the HL-LHC.
In this likelihood scan, only $c_\varphi$ is varied and all other SMEFT coefficients are set to zero.
We show the likelihoods for the differential analyses of $hh\to b\bar{b}\gamma\gamma$ and $hh\to b\bar{b}\tau^+\tau^-$ and for the inclusive signal strength $\mu_{hh}$, in all cases corresponding to an integrated luminosity of $\mathcal{L}=3$ ab$^{-1}$ (hence, adding up to a total of $\mathcal{L}=6$ ab$^{-1}$ as corresponding to the ATLAS+CMS combination).
The likelihood scan of $\mu_{hh}$ displays the expected second solution for $\kappa_\lambda\ne 1$, which is lifted by adding the information on $hh$ production differential in $m_{hh}$ in the two dominant final states.
The combined HL-LHC likelihood is Gaussian  to a good approximation. 
We have verified that by applying the $\Delta\chi^2=1$ criterion we approximately reproduce the projected precision for $\kappa_\lambda$ extracted from the exact HL-LHC likelihood of the ATLAS+CMS combination, namely $\delta \kappa_\lambda=_{-26\%}^{+29\%}$.

\section{Overview of numerical bounds on $c_\varphi$ and $\delta \kappa_3$}
\label{app:likelihood}

Here, we collect the projected bounds on $c_\varphi$ and $\delta\kappa_\lambda$ in Table~\ref{tab:bounds_all}. These correspond to the bounds shown in Fig.~\ref{fig:bound_plot_cphi_hllhc_fcc}.

\begin{table}[h!]
    \centering
    \renewcommand{\arraystretch}{1.24}
    \begin{tabular}{|l | c|c|c|c|c|}
    \hline
         \multicolumn{2}{|l|}{\multirow{2}{*}{$c_{\varphi}$ [TeV$^{-2}$]
         ~($\mu_0=250$ GeV)}} & \multicolumn{2}{c|}{Individual} & \multicolumn{2}{c|}{Marginalised} \\
         \cline{3-6}
          \multicolumn{2}{|l|}{} & $68\%$ C.I. & $95\%$ C.I. & $68\%$ C.I. & $95\%$ C.I. \\
         \hline
         \multirow{2}{*}{LHC Run 2} & $\quad$ Linear $\quad$ & $\quad$$[-3.51,\,3.51]$$\quad$ & $\quad$$[-6.88,\,6.88]$$\quad$ &$\quad$ $[-3.57,\,3.57]$$\quad$ & $\quad$$[-7.06,\,7.06]$$\quad$ \\
         \cline{2-6}
                  & Quad & $\quad$$[-7.55,\,0.76]$$\quad$ & $[-9.57,\,3.53]$ & $[-7.38,\,0.75]$ & $[-9.69,\,3.16]$ \\
         \hline
                  \multirow{2}{*}{HL-LHC } & Linear & $\quad$$[- 0.60,\,0.60]$$\quad$ & $[-1.17,\,1.17]$ & $[-0.69,\,0.69]$ & $[-1.35,\,1.35 ]$ \\
         \cline{2-6}
                  & Quad & $[- 0.74,\,0.62]$ & $[- 2.38,\,1.31]$ & $[- 0.94,\,0.65]$ & $[-2.95,\,1.24]$ \\
         \hline
         \multirow{2}{*}{HL-LHC\,\&\,FCC-ee@240 } & Linear & $[ - 0.19,\,0.19]$ & $[- 0.37,\,0.37]$ & $[- 0.63,\,0.63]$ & $[-1.25,\,1.25]$ \\
         \cline{2-6}
         \multirow{1}{*}{}& Quad. & $[-0.18,\,0.19]$ & $[-0.37,\,0.36]$ & $[-0.89,\,0.44]$ & $[-1.82,\,0.98]$ \\
         \hline
         \multirow{2}{*}{HL-LHC\,\&\,FCC-ee } & Linear & $[- 0.18,\,0.18]$ & $[-0.37,\,0.37]$ & $[-0.38,\,0.38]$ & $[-0.73,\,0.73]$ \\
         \cline{2-6}
         \multirow{1}{*}{ }& Quad. & $[-0.19,\,0.18]$ & $[-0.37,\,0.37]$ & $[- 0.36,\,0.32]$ & $[-0.72,\,0.64]$ \\
         \hline
    \end{tabular}
    \\[0.5cm]
    \begin{tabular}{|l | c|c|c|c|c|}
    \hline
         \multicolumn{2}{|l|}{\multirow{2}{*}{$\delta\kappa_3$~($\mu_0=250$ GeV)}} & \multicolumn{2}{c|}{Individual} & \multicolumn{2}{c|}{Marginalised} \\
         \cline{3-6}
          \multicolumn{2}{|l|}{} & $68\%$ C.I. & $95\%$ C.I. & $68\%$ C.I. & $95\%$ C.I. \\
         \hline
         \multirow{2}{*}{LHC Run 2} & $\quad$ Linear $\quad$ & $\quad$$[-1.65,\,1.65]$$\quad$ & $\quad$$[-3.24,\,3.24]$$\quad$ & $\quad$$[-1.68,\,1.68]$ $\quad$& $\quad$$[-3.34,\,3.34]$$\quad$ \\
         \cline{2-6}
                  & Quad. & $[-0.36,\,3.55]$ & $[-1.66,\,4.50]$ & $[-0.54,\,3.27]$ & $[-1.49,\,4.56]$ \\
         \hline
                  \multirow{2}{*}{HL-LHC } & Linear & $[-0.28,\,0.28]$ & $[-0.55,\,0.55]$ & $[-0.32,\,0.32]$ & $[-0.64,\,0.64]$ \\
         \cline{2-6}
                  & Quad & $[- 0.29,\,0.35]$ & $[-0.62,\,1.12]$ & $[-0.26,\,0.49]$ & $[-0.66,\,1.30]$ \\
         \hline
         \multirow{2}{*}{HL-LHC\,\&\,FCC-ee@240 } & Linear & $[-0.09,\,0.09]$ & $[-0.17,\,0.17]$ & $[-0.29,\,0.29]$ & $[-0.58,\,0.58]$ \\
         \cline{2-6}
         \multirow{1}{*}{ }& Quad & $[-0.09,\,0.09]$ & $[-0.17,\,0.17]$ & $[-0.20,\,0.41]$ & $[-0.45,\,0.83]$ \\
         \hline
         \multirow{2}{*}{HL-LHC\,\&\,FCC-ee } & Linear & $[-0.09,\,0.09]$ & $[-0.17,\,0.17]$ & $[-0.17,\,0.17]$ & $[-0.34,\,0.34]$ \\
         \cline{2-6}
         \multirow{1}{*}{ }& Quad & $[-0.09,\,0.09]$ & $[-0.17,\,0.17]$ & $[-0.15,\,0.16]$ & $[-0.29,\,0.33]$  \\
         \hline
    \end{tabular}
    \caption{$68\%$ and $95\%$ C.I. for  $c_\varphi$ (upper table) and $\delta\kappa_3$ (lower table) at $\mu_0=250$ GeV, corresponding to the results shown in Table~\ref{tab:bounds_maintext}.
Individual and global marginalised fits are compared for different input datasets.
    \label{tab:bounds_all}
}
\end{table}

\end{document}